\documentclass[12pt,twoside]{article}

\usepackage{pslatex}	
\usepackage{amsfonts}
\usepackage{amsmath}
\usepackage{amsthm}
\usepackage{amscd}
\usepackage{cite}
\usepackage{epsf}
\usepackage{a4}

\def\beq{\begin{equation}}
\def\eeq{\end{equation}}
\def\bea{\begin{eqnarray}}
\def\eea{\end{eqnarray}}

\def\beann{\begin{eqnarray*}}
\def\eeann{\end{eqnarray*}}

\let\a=\alpha \let\be=\beta \let\g=\gamma 
  \let\h=\eta 

  \let\la=\lambda \let\m=\mu
 \let\x=\xi \let\p=\pi \let\r=\rho \let\s=\sigma
\let\om=\omega 
\let\ph=\varphi   
  \let\Th=\Theta
\let\La=\Lambda  \let\D=\Delta

\let\qd=\quad  

\def\epp{\, .}
\def\epc{\, ,}

\def\tst#1{{\textstyle #1}}

\theoremstyle{plain}

\newtheorem{lemma}{Lemma}

\newtheorem*{corollary*}{Corollary}

\theoremstyle{definition}

\def\2{\frac{1}{2}} \def\4{\frac{1}{4}}

\def\6{\partial}

\def\+{\dagger}

\def\<{\langle} \def\>{\rangle}

\let\auf=\uparrow \let\ab=\downarrow

\renewcommand\i{{\rm i}}

\def\re{{\rm e}}

\DeclareMathOperator{\sh}{sh}
\DeclareMathOperator{\ch}{ch}

\def\tr{{\rm tr}}

\def\fa{\mathfrak{a}}
\def\faq{\overline{\mathfrak{a}}}

\def\aqq{\widetilde{\alpha}}
\def\bqq{\widetilde{\beta}}

\renewcommand{\appendix}{%
   \renewcommand{\section}{
	\secdef\Appendix\sAppendix}%
   \setcounter{section}{0}%
   \renewcommand{\thesection}{\Alph{section}}%
   \renewcommand{\theequation}{\thesection.\arabic{equation}}%
}

\newcommand{\Appendix}[2][?]{%
     \refstepcounter{section}%
     \setcounter{equation}{0}%
     \addcontentsline{toc}{appendix}%
          {\protect\numberline{\appendixname~\thesection} #1}%
     \vspace{\baselineskip}%
     {\noindent\large\bfseries\appendixname: #2\par}%
     \sectionmark{#1}\vspace{\baselineskip}}

\newcommand{\sAppendix}[1]{%
     {\noindent\large\bfseries\appendixname\:: #1\par}%
     \sectionmark{#1}\vspace{\baselineskip}}

\begin{document}

\thispagestyle{empty}

\begin{center}

{\Large {\bf Integral representation of the density matrix of the XXZ
chain at finite temperatures\\}}

\vspace{7mm}

{\large Frank G\"{o}hmann,\footnote[2]{e-mail:
goehmann@physik.uni-wuppertal.de} Andreas Kl\"{u}mper\footnote[1]{%
e-mail: kluemper@physik.uni-wuppertal.de} and Alexander Seel%
\footnote[3]{e-mail: seel@physik.uni-wuppertal.de}\\

\vspace{5mm}

Fachbereich C -- Physik, Bergische Universit\"at Wuppertal,\\
42097 Wuppertal, Germany\\}

\vspace{20mm}

{\large {\bf Abstract}}

\end{center}

\begin{list}{}{\addtolength{\rightmargin}{10mm}
               \addtolength{\topsep}{-5mm}}
\item
We present an integral formula for the density matrix of a finite
segment of the infinitely long spin-$\2$ XXZ chain. This formula is
valid for any temperature and any longitudinal magnetic field.
\\[2ex]
\end{list}

\clearpage

\section{Introduction}
A system which is part of a larger system and interacts with its
other parts cannot be in a pure quantum state and must be described
by a density matrix \cite{LaLiIII}. The calculation of the density
matrix involves taking the trace over all those degrees of freedom of
the larger system which do not belong to the subsystem we are
interested in and, in the thermodynamic limit, when the subsystem stays
finite, but the large system becomes infinitely large, is usually rather
hard. One of the few examples where a density matrix could be
calculated for an interacting system is the antiferromagnetic spin-$\2$
XXZ chain with Hamiltonian
\begin{equation} \label{xxzham}
     H_{XXZ} = J \sum_{j=1}^L \Bigl( \s_{j-1}^x \s_j^x
                  + \s_{j-1}^y \s_j^y + \D (\s_{j-1}^z \s_j^z - 1)
		     \Bigr)
\end{equation}
acting on the tensor product of spaces of states of $L$ spins $\2$.
The $\s^\a$, $\a = x, y, z$, in (\ref{xxzham}) are the Pauli matrices
and $J > 0$ and $\D > - 1$ are two real parameters, the exchange
interaction and the exchange anisotropy.

In \cite{JMMN92,JiMi96,KMT99b} integral formulae for the zero
temperature density matrix elements of a segment of length $m$ of the
infinite chain were obtained. Since the expectation value of any
operator acting on a segment of length $m$ can be expressed in terms
of the density matrix elements (see equation (\ref{cordensmat}) below),
the density matrix enables one, in particular, to calculate the
correlations of local observables (for recent developments see
\cite{KMST04cpp}). This is the reason why the density matrix elements
were called `elementary blocks of correlation functions' in
\cite{JMMN92,JiMi96,KMT99b} and also is the actual reason why we got
interested in the subject. In fact, much recent progress in the
calculation of short range correlations for the XXZ chain
\cite{BoKo01,BoKo02,BST04,KSTS03,KSTS04,SSNT03,TKS04} originates in
the integral representation of the density matrix obtained in
\cite{JMMN92,JiMi96,KMT99b}.

Below we generalize the formulae first obtained in \cite{JMMN92,JiMi96,%
KMT99b} to finite temperatures. It turns out that the same means that
where successfully applied in the calculation of a generating function
of the $S^z$-$S^z$ correlation functions at finite temperatures in
\cite{GKS04a} also work for the density matrix. Namely we may
combine algebraic Bethe ansatz techniques for the calculation of matrix
elements \cite{Korepin82,Slavnov89,KBIBo,KMT99b} with the quantum
transfer matrix approach to the thermodynamics of quantum spin chains
\cite{Suzuki85,SuIn87,Kluemper92,Kluemper93}.

\section{The quantum transfer matrix}
All properties of the XXZ chain can be derived from the well-known
trigonometric solution
\begin{equation} \label{rxxz}
     R(\la,\m) = \begin{pmatrix}
                    1 & 0 & 0 & 0 \\
		    0 & b(\la, \m) & c(\la, \m) & 0 \\
		    0 & c(\la, \m) & b(\la, \m) & 0 \\
		    0 & 0 & 0 & 1
		 \end{pmatrix}
\end{equation}
of the Yang-Baxter equation, where
\begin{equation} \label{defbc}
     b(\la, \m) = \frac{\sh(\la - \m)}{\sh(\la - \m + \h)} \epc \qd
     c(\la, \m) = \frac{\sh(\h)}{\sh(\la - \m + \h)} \epp
\end{equation}
This $R$-matrix not only generates the Hamiltonian (\ref{xxzham}),
\begin{equation} \label{fundhdens}
     H_{XXZ} = 2 J \sh(\h) \sum_{j=1}^L
               \6_\la (P R)_{j-1, j} (\la,0) \Big|_{\la = 0}
\end{equation}
($P$ is the permutation matrix, $\ch(\h) = \D$ in (\ref{xxzham})), but
also a related auxiliary vertex model whose partition function in a
certain `Trotter limit' is equal to the partition function of the XXZ
Hamiltonian. Details of this construction in a notation also suitable
for our present purpose were reviewed in our paper \cite{GKS04a}. Here
we shall repeat only the most important formulae as far as they are
needed to understand the notation below. We define the monodromy
matrix of the auxiliary vertex model as
\begin{equation} \label{monoqtm}
     T_j (\la ) = \begin{pmatrix} A(\la) & B(\la) \\ C(\la) & D(\la)
                  \end{pmatrix}_j =
        R_{j \overline{N}} \bigl(\la, \tst{\frac{\be}{N}} \bigr)
	R_{\overline{N-1} j}^{t_1}
	   \bigl(- \tst{\frac{\be}{N}}, \la \bigr) \dots
        R_{j \overline{2}} \bigl(\la, \tst{\frac{\be}{N}} \bigr)
	R_{\bar 1 j}^{t_1} \bigl(- \tst{\frac{\be}{N}}, \la \bigr) \epc
\end{equation}
where $t_1$ means transposition with respect to the first space, and
the parameter $\be$ is inversely proportional to the temperature $T$,
\begin{equation}
     \be = \frac{2J \sh(\h)}{T} \epp
\end{equation}
The monodromy matrix (\ref{monoqtm}) acts in the tensor product of an
auxiliary space $j$ and $N$ `quantum spaces' $\bar 1, \dots,
\overline{N}$. It generates a representation of the Yang-Baxter algebra
with $R$-matrix (\ref{rxxz}). The corresponding transfer matrix
\begin{equation} \label{tqtm}
     t(\la) = \tr_j T_j (\la) \epc
\end{equation}
called the quantum transfer matrix, defines a vertex model, whose
partition function in the Trotter limit $N \rightarrow \infty$ is equal
to the partition function $Z_L = \tr \exp (- H_{xxz}/T)$ of the XXZ
chain of length $L$,
\begin{equation} \label{finitepart}
     Z_L = \lim_{N \rightarrow \infty} \tr_{\bar 1 \dots \overline{N}}
	      \bigl( t^{QTM} (0) \bigr)^L
         = \sum_{j=0}^\infty \La^L_n (0) \epp
\end{equation}
Here $\La_n (\la)$ denotes the $n$th eigenvalue of the quantum
transfer matrix. Note that there is a unique real eigenvalue
$\La_0 (\la)$ with largest modulus which dominates the partition
function in the thermodynamic limit, when $L$ goes to infinity.
This single leading eigenvalue determines the bulk thermodynamics of
the XXZ chain. We showed in \cite{GKS04a} that the corresponding
eigenvector determines the state of thermodynamic equilibrium
completely. It fixes all finite temperature correlation functions
(compare \cite{Suzuki03a,Suzuki03b}).

Due to the conservation of the $z$-component of the total spin
\begin{equation}
     S^z = \tst{\2} \sum_{j = 1}^L \s_j^z
\end{equation}
thermodynamics and finite temperature correlation functions can still
be treated within the quantum transfer matrix approach if the system
is exposed to an external magnetic field $h$ in $z$-direction
\cite{GKS04a}. The external field is properly taken into account by
applying a twist to the monodromy matrix (\ref{monoqtm}),
\begin{equation}
     T (\la) \rightarrow
        T (\la) \Bigl( \begin{smallmatrix} \re^{h/2T} & 0 \\
	               0 & \re^{- h/2T} \end{smallmatrix} \Bigr) \epp
\end{equation}

\section{Integral formula for density matrix elements}
\label{sec:result}
We shall use the notation
\begin{equation}
     e_1^1 = \tst{\begin{pmatrix} 1 & 0 \\ 0 & 0 \end{pmatrix}} \epc \qd
     e_1^2 = \begin{pmatrix} 0 & 1 \\ 0 & 0 \end{pmatrix} \epc \qd
     e_2^1 = \begin{pmatrix} 0 & 0 \\ 1 & 0 \end{pmatrix} \epc \qd
     e_2^2 = \begin{pmatrix} 0 & 0 \\ 0 & 1 \end{pmatrix}
\end{equation}
for the gl(2) standard basis. The canonical embedding of these matrices
into the space of operators on the space of states 
$({\mathbb C}^2)^{\otimes L}$ of the Hamiltonian (\ref{xxzham}) will
be denoted by ${e_j}^\a_\be$, $\a, \be = 1, 2$, $j = 1, \dots, L$.
Then every operator $A_{1, \dots, m}$ that acts on sites $1$ to $m$ of
the spin chain can be expanded as
\begin{equation}
     A_{1, \dots, m} = A^{\be_1 \dots \be_m}_{\a_1 \dots \a_m}
                       {e_1}_{\be_1}^{\a_1} \dots {e_m}_{\be_m}^{\a_m}
		       \epc
\end{equation}
where implicit summation over the Greek indices is implied. The thermal
average of such type of operator is
\begin{equation} \label{cordensmat}
     \< A_{1, \dots, m} \>_{T, h} 
        = A^{\be_1 \dots \be_m}_{\a_1 \dots \a_m}
	  \bigl\< {e_1}_{\be_1}^{\a_1}
	           \dots {e_m}_{\be_m}^{\a_m} \bigr\>_{T, h} \epp
\end{equation}
Thus, it is sufficient to calculate the expectation values
$\< {e_1}_{\be_1}^{\a_1} \dots {e_m}_{\be_m}^{\a_m} \bigr\>_{T, h}$
which define the matrix elements of the density matrix of a chain
segment of length~$m$.

Following \cite{GKS04a} we can calculate the general density matrix
element as a limit of an appropriately defined inhomogeneous finite
Trotter number approximant,
\begin{equation} \label{defdensmat}
     \bigl\<{e_1}_{\be_1}^{\a_1} \dots
            {e_m}_{\be_m}^{\a_m}\bigr\>_{T, h} =
        \lim_{N \rightarrow \infty} \:
	\lim_{\x_1, \dots, \x_m \rightarrow 0}
        {D_N}^{\a_1 \dots \a_m}_{\be_1 \dots \be_m} (\x_1, \dots, \x_m)
	\epc
\end{equation}
where
\begin{equation} \label{defdn}
     {D_N}^{\a_1 \dots \a_m}_{\be_1 \dots \be_m} (\x_1, \dots, \x_m) =
	\frac{\< \{\la\}| T^{\a_1}_{\be_1} (\x_1)
	                     \dots T^{\a_m}_{\be_m} (\x_m)|\{\la\}\>}
             {\<\{\la\}| \prod_{j=1}^m t (\x_j) |\{\la\}\>}
\end{equation}
for $\a_j, \be_k = 1, 2$, and $|\{\la\}\> = B(\la_1) \dots B(\la_{N/2})
|0\>$ is the eigenstate corresponding to the leading eigenvalue $\La_0
(\la)$ which is parameterized by a specific set of Bethe roots
$\{\la\} = \{\la_j\}_{j=1}^{N/2}$. The complex inhomogeneity parameters
$\x_j$ `regularize' the expression in the numerator on the right hand
side of (\ref{defdn}). Moreover, at least in the zero temperature case,
it turned out to be useful and interesting to study the dependence of
$D_N$ on these parameters \cite{BKS03,BJMST04}.

We may identify values 1 of the indices with the symbol $\auf$,
representing an up-spin, and values 2 with $\ab$, representing
a down spin. Then the upper and lower indices $(\a_n)_{n=1}^m$ and
$(\be_n)_{n=1}^m$ in e.g.\ (\ref{defdn}) may be
visualized as sequences of up- and down-spins. We shall denote the
position $n$ of the $j$th up-spin in the sequence $(\a_n)_{n=1}^m$ by
$\a_j^+$ and the number of up-spins in $(\a_n)_{n=1}^m$ by $|\a^+|$.
Similarly we define $\be_j^-$ as the position of the $j$th down-spin
in $(\be_n)_{n=1}^m$ and denote the number of down-spins in
$(\be_n)_{n=1}^m$ by $|\be^-|$. This yields two sequences
$(\a_j^+)_{j=1}^{|\a^+|}$ and $(\be_j^-)_{j=1}^{|\be^-|}$ which
\renewcommand{\arraystretch}{1.5}
\tabcolsep3mm
\begin{table}

{
\large
\begin{center}
\begin{tabular}{c||cccccccc}
$j$ & 1 & 2 & 3 & 4 & 5 & 6 & 7 & 8 \\ \hline \hline
$\a_j$ & $\ab$ & $\ab$ & $\auf$ & $\auf$ & $\ab$ & $\auf$ & $\ab$ &
$\ab$ \\
$\be_j$ & $\ab$ & $\auf$ & $\ab$ & $\ab$ & $\ab$ & $\auf$ & $\auf$ &
$\ab$ \\ \hline
$\a_j^+$ & 3 & 4 & 6 &&&&& \\
$\be_j^-$ & 1 & 3 & 4 & 5 & 8 &&& \\ \hline
$\aqq_j^+$ & 6 & 4 & 3 &&&&& \\
$\bqq_j^-$ &&&& 1 & 3 & 4 & 5 & 8 \\ \hline
$\g_j$ & 1 & 3 & 3 & 4 & 4 & 5 & 6 & 8 \\ \hline
$\g_j^+$ & 3 & 5 & 7 &&&&& \\ \hline
$\g_j^-$ & 1 & 2 & 4 & 6 & 8 &&&
\end{tabular}
\end{center}
}

\caption{\label{tab:ex1} Example for the definition of the sequences
$(\a_j^+)$, $(\be_j^-)$, $(\aqq_j^+)$ and $(\bqq_j^-)$. The pattern
corresponds to the string $D(\x_1) C(\x_2) B(\x_3) B(\x_4) D(\x_5)
A(\x_6) C(\x_7) D(\x_8)$ of monodromy matrix elements, m = 8,
$|\a^+| = 3$, $|\be^-| = 5$. The sequences $(\g_j)$ and $(\g_j^\pm)$
are defined in section \ref{sec:proof} below where they are needed.}
\end{table}
we reorder and shift by the prescription $\aqq_j^+ =
\a_{|\a^+| - j +1}^+$, $j = 1, \dots, |\a^+|$, and $\bqq_j^- =
\be_{j - |\a^+|}^-$, $j = |\a^+| + 1, \dots, |\a^+| + |\be^-|$. The
definitions are illustrated with an example in table \ref{tab:ex1}.

Due to the conservation of the total spin the matrix elements
${D_N}^{\a_1 \dots \a_m}_{\be_1 \dots \be_m} (\x_1, \dots, \x_m)$
vanish if $|\a^+| + |\be^-| \ne m$. They are non-trivial only if
$|\a^+| + |\be^-| = m$. For this case we suggest the integral
representation
\begin{align} \label{densint}
     {D_N}^{\a_1 \dots \a_m}_{\be_1 \dots \be_m} (\x_1, \dots, & \x_m)
        = \notag \\
	  \prod_{j=1}^{|\a^+|}
	     \int_{\cal C} & \frac{d \om_j}{2 \p \i (1 + \fa (\om_j))}
	     \prod_{k=1}^{\aqq_j^+ - 1} \sh(\om_j - \x_k - \h)
	     \prod_{k = \aqq_j^+ + 1}^m \sh(\om_j - \x_k) \notag \\
          \prod_{j = |\a^+| + 1}^{m}
	     \int_{\cal C} & \frac{d \om_j}{2 \p \i (1 + \faq (\om_j))}
	     \prod_{k=1}^{\bqq_j^- - 1} \sh(\om_j - \x_k + \h)
	     \prod_{k = \bqq_j^- + 1}^m \sh(\om_j - \x_k) \notag \\ &
        \frac{\det( - G(\om_j, \x_k))}
	     {\prod_{1 \le j < k \le m}
	         \sh(\x_k - \x_j) \sh( \om_j - \om_k - \h)} \epp
\end{align}
Equation (\ref{densint}) holds for any finite Trotter number $N$.
The Trotter number enters the right hand side of (\ref{densint})
implicitly through the functions $\fa (\om)$, $\faq (\om) = 1/\fa (\om)$
and $G(\om, \x)$. Performing the Trotter limit for $D_N$ means to
replace these functions by their respective Trotter limits. For brevity
we show only the non-linear integral equation that determines the
Trotter limit of $\fa$,
\begin{equation} \label{nlieh}
     \ln \fa (\la) = - \frac{h}{T}
                     - \frac{2J \sh^2 (\h)}{T \sh(\la) \sh(\la + \h)}
                     - \int_{\cal C} \frac{d \om}{2 \p \i} \,
	               \frac{\sh (2 \h) \ln (1 + \fa (\om))}
		            {\sh(\la - \om + \h)
			            \sh(\la - \om - \h)} \epp
\end{equation}
The corresponding finite Trotter number equation can be found in
\cite{GKS04a}. The function $G(\om, \x)$ which was introduced in
\cite{GKS04a} has to be calculated from the linear integral
\begin{figure}

\begin{center}

\epsfxsize \textwidth
\epsffile{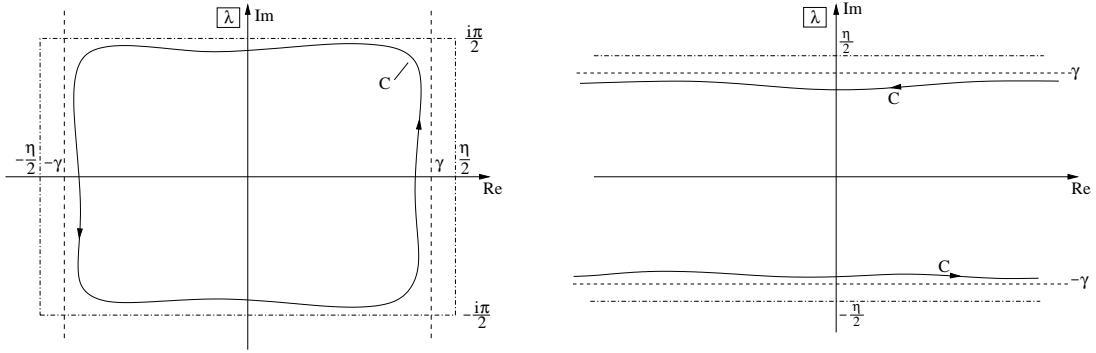}

\caption{\label{fig:cancon} The canonical contour ${\cal C}$ in the
off-critical regime $\D > 1$ (left panel) and in the critical regime
$|\D | < 1$ (right panel). For $\D > 1$ the contour is a rectangle
with sides at $\pm \i \frac{\p}{2}$ and $\pm \g$, where $\g$ is
slightly smaller than $\frac{\h}{2}$. For $|\D| < 1$ the contour
surrounds the real axis at a distance $|\g|$ slightly less than
$\frac{|\h|}{2}$.}
\end{center}
\end{figure}
equation
\begin{multline} \label{defg}
     G(\la,\x) = \frac{\sh(\h)}{\sh(\x - \la) \sh(\x - \la + \h)} \\
               + \int_{\cal C} \frac{d \om}{2 \p \i (1 + \fa (\om))} \,
	         \frac{\sh (2 \h)G (\om,\x)}
		      {\sh(\la - \om + \h)\sh(\la - \om - \h)}
\end{multline}
and generalizes the `density function' $\r (\la)$, that determines 
the ground state energy and the zero temperature magnetization,
to finite temperatures and to the inhomogeneous case \cite{GKS04a}.
The canonical contour ${\cal C}$ in (\ref{densint})-(\ref{defg})
depends on $\h$ and is shown in figure~\ref{fig:cancon}.

Our conjecture (\ref{densint}) is based on the following observations.
(i) Equation (\ref{densint}) is true for $m = 1, 2$ which can easily
be verified using the commutation relations comprised in the
Yang-Baxter algebra and equations (98) and (104) of \cite{GKS04a}.
(ii) In two cases equation (\ref{densint}) reduces to the approximant
to the emptiness formation probability, namely for $\a_j = \be_j = 1$
for $j = 1, \dots, m$ and $\a_j = \be_j = 2$ for $j = 1, \dots, m$.
These two cases correspond to taking the limits $\ph \rightarrow \pm
\infty$ in our formulae for the generating function of the $S^z$-$S^z$
correlation functions in \cite{GKS04a}. One arrives at (\ref{densint})
by applying the result of appendix C of \cite{KMST02a} (for more
details see \cite{GKS04bpp}). (iii) In the zero temperature limit
(\ref{densint}) reduces to the result of \cite{KMT99b} which, in turns,
generalizes the formulae \cite{JMMN92,JiMi96} of Jimbo et al.\ to
finite values of the external magnetic field. Some important
intermediate steps of a proof of (\ref{densint}) in the general case
are sketched in the next section.

Taking the Trotter limit and the homogeneous limit (for the latter
compare \cite{KMT99b}) in (\ref{densint}) we obtain an integral
formula for the density matrix elements (\ref{defdensmat}),
\begin{align} \label{densinthom}
     \bigl\<{e_1}_{\be_1}^{\a_1} \dots
            {e_m}_{\be_m}^{\a_m}\bigr\>_{T, h} =
        \prod_{j=1}^{|\a^+|}
	     \int_{\cal C} & \frac{d \om_j}{2 \p \i (1 + \fa (\om_j))}
	     \sh^{\aqq_j^+ - 1} (\om_j - \h)
	     \sh^{m - \aqq_j^+} (\om_j) \notag \\
        \prod_{j = |\a^+| + 1}^{m}
	     \int_{\cal C} & \frac{d \om_j}{2 \p \i (1 + \faq (\om_j))}
	     \sh^{\bqq_j^- - 1} (\om_j + \h)
	     \sh^{m - \bqq_j^-} (\om_j) \notag \\ & \mspace{-36.0mu}
	\det \biggl[
	   - \frac{\6^{(k-1)}_\x G(\om_j, \x)|_{\x = 0}}{(k - 1)!}
	     \biggr]
        \frac{1}{\prod_{1 \le j < k \le m} \sh( \om_j - \om_k - \h)}
	\epp
\end{align}
We should mention that the functions $1/(1 + \fa(\om))$ and
$1/(1 + \faq(\om))$, respectively, appear quite naturally here, since
they generalize the Fermi functions for particles and holes to the
interacting case.

\section{Elements of a proof of the integral formula}
\label{sec:proof}
By analogy with the example of the generating function of the
$S^z$-$S^z$ correlation functions treated in \cite{GKS04a} we expect
that a proof of the integral formula (\ref{densint}) for the general
density matrix element can be achieved in three steps. Step 1 consists
in calculating the action of a string of operators $T^{\a_1}_{\be_1}
(\x_1) \dots T^{\a_m}_{\be_m} (\x_m)$ on the left Bethe state
$\<\{\la\}| = \<0| C(\la_{N/2}) \dots C(\la_1)$. As it is clear from the
structure of the Yang-Baxter algebra, the result is a linear combination
of vectors of the same form with some of the $\la_j$ replaced with
inhomogeneity parameters $\x_k$. Step 2 is to calculate the scalar
product of the vectors occurring in the sum of step 1 with the
Bethe state $|\{\la\}\>$, to divide by the norm and by the product
$\prod_{j=1}^m \La_0 (\x_j)$ of transfer matrix eigenvalues, and to
rewrite the resulting expression in a form that is suitable for
taking the Trotter limit. This step is the same as in the derivation
of the integral formula for the generating function of the $S^z$-$S^z$
correlation function in \cite{GKS04a}. Thus, we can use our former
results.
\begin{lemma}
\cite{GKS04a}
\begin{multline} \label{ratio}
     \frac{\<\{\x^+\} \cup \{\la^-\}|\{\la\}\>}
          {\<\{\la\}|\{\la\}\>\prod_{j=1}^m \La_0 (\x_j)}
        = \biggl[ \prod_{j = 1}^{|\la^-|} \prod_{k=1}^{|\la^+|}
	          b(\la_j^-, \la_k^+) \biggr] \\
	  \biggl[ \prod_{j=1}^{|\x^-|}
	          \frac{1}{a(\x_j^-)(1 + \fa (\x_j^-))}
	          \prod_{k = 1}^{|\la^-|} b(\la_k^-, \x_j^-) \biggr]
          \biggl[ \prod_{j=1}^{|\la^+|}
	          \frac{1}{a(\la_j^+) \fa' (\la_j^+)}
	          \prod_{k=1}^m b(\la_j^+, \x_k) \biggr] \\
	  \biggl[ \prod_{j,k=1}^{|\x^+|}
	         \frac{\sh(\la_j^+ - \x_k^+ + \h)}
	              {\sh(\la_j^+ - \la_k^+ + \h)} \biggr]
	  \biggl[ \prod_{1 \le j < k \le |\x^+|}
	         \frac{\sh(\la_j^+ - \la_k^+)}{\sh(\x_j^+ - \x_k^+)}
		 \biggr]
	  \det G(\la_j^+,\x_k^+) \epp
\end{multline}
Here we divided the sets $\{\la\}$ and $\{\x\}$ into disjoint subsets,
$\{\la^\pm\}$ and $\{\x^\pm\}$, and employed the notation $|X|$ for the
number of elements in a set $\{X\}$. The function $a(\la)$ on the right
hand side is the vacuum eigenvalue of the monodromy matrix element
$A(\la)$.
\end{lemma}
In step 3 we have to transform the sums obtained in steps 1 and 2 into
integrals over the canonical contour ${\cal C}$. This involves the
calculation of residua and the resummation of the many terms emerging
in this procedure.

So far we could not complete step 3 in the general case, but only for
examples of small $m$. Yet, we have obtained a complete and
satisfactory result for step 1 that can be summarized in the following
\begin{lemma} \label{lem:genact}
Multiple action of monodromy matrix elements on a general state. Let
$\la_1, \dots, \la_{M + m} \in {\mathbb C}$ be mutually distinct. Then
\begin{multline} \label{genact}
     \<0| \Bigl[ \prod_{k = 1}^M C(\la_k) \Bigr]
        T^{\a_1}_{\be_1} (\la_{M+1}) \dots T^{\a_m}_{\be_m} (\la_{M+m})
	= \\
	  \sum_{\ell_1 = 1}^{M + \g_1}
	  \sum_{\substack{\ell_2 = 1 \\ \ell_2 \ne \ell_1}}^{M + \g_2}
	  \dots
	  \sum_{\substack{\ell_m = 1 \\ \ell_m \ne \ell_1, \dots,
	                  \ell_{m-1}}}^{M + \g_m}
	  \biggl[ \prod_{j=1}^{|\a^+|} a(\la_{\ell_{\g_j^+}})
	          c(\la_{M + \a_j^+}, \la_{\ell_{\g_j^+}})
	          \prod_{\substack{k = 1 \\ k \ne \ell_1, \dots,
	 	                   \ell_{\g_j^+}}}^{M + \a_j^+}
		  \frac{1}{b(\la_k, \la_{\ell_{\g_j^+}})} \biggr] \\
	  \biggl[ \prod_{j=1}^{m-|\a^+|} d(\la_{\ell_{\g_j^-}})
	          c(\la_{\ell_{\g_j^-}}, \la_{M + \be_j^-})
	          \prod_{\substack{k = 1 \\ k \ne \ell_1, \dots,
		                   \ell_{\g_j^-}}}^{M + \be_j^-}
		  \frac{1}{b(\la_{\ell_{\g_j^-}}, \la_k)} \biggr]
          \<0| \Bigl[ \mspace{-9.0mu}
	       \prod_{\substack{k = 1 \\ k \ne \ell_1, \dots,
	                        \ell_m}}^{M + m}
	       \mspace{-18.0mu} C(\la_k) \Bigr] \epp
\end{multline}
Here $a(\la)$ and $d(\la)$ are the vacuum eigenvalues of $A(\la)$ and
$D(\la)$, respectively. The sequences $(\a_j^+)$ and $(\be_k^-)$ were
defined in section \ref{sec:result} above. We arrange all $\a_j^+$ and
$\be_k^-$ in non-decreasing order, in such a way that $\be_k^-$
appears left to $\a_j^+$ if $\be_k^- = \a_j^+$. This defines the
sequence $(\g_n)_{n=1}^m$. The position of $\a_j^+$ in this sequence
is denoted by $\g_j^+$ and the position of $\be_k^-$ by $\g_k^-$
(see table \ref{tab:ex1} for an example).
\end{lemma}
Lemma \ref{lem:genact} can be proven by induction over $m$. Using the
fact that $c(\la,\la) = 1$, the well-known \cite{KBIBo} `elementary'
commutation relations for moving $A(\la)$, $B(\la)$ or $D(\la)$ through
a product of $C$s can be rewritten in the form
\begin{align}
     \<0| & \Bigl[ \prod_{k=1}^M C(\la_k) \Bigr] A (\la_{M+1})
        = \sum_{\ell = 1}^{M+1} a(\la_\ell) c(\la_{M+1}, \la_\ell)
	  \Bigr[ \prod_{\substack{k = 1 \\ k \ne \ell}}^{M+1}
	         \frac{1}{b(\la_k, \la_\ell)} \Bigl]
		 \<0| \prod_{\substack{k = 1 \\ k \ne \ell}}^{M+1}
		 C(\la_k) \epc \\
     \<0| & \Bigl[ \prod_{k=1}^M C(\la_k) \Bigr] D (\la_{M+1})
        = \sum_{\ell = 1}^{M+1} d(\la_\ell) c(\la_\ell, \la_{M+1})
	  \Bigr[ \prod_{\substack{k = 1 \\ k \ne \ell}}^{M+1}
	         \frac{1}{b(\la_\ell, \la_k)} \Bigl]
		 \<0| \prod_{\substack{k = 1 \\ k \ne \ell}}^{M+1}
		 C(\la_k), \\
     \<0| & \Bigl[ \prod_{k=1}^M C(\la_k) \Bigr] B (\la_{M+1})
        = \sum_{\ell_1 = 1}^{M+1}
	  \sum_{\substack{\ell_2 = 1 \\ \ell_2 \ne \ell_1}}^{M+1}
	  d(\la_{\ell_1}) c(\la_{\ell_1}, \la_{M+1})
	  \Bigr[ \prod_{\substack{k = 1 \\ k \ne \ell_1}}^{M+1}
	         \frac{1}{b(\la_{\ell_1}, \la_k)} \Bigl]
		 \notag \\ & \mspace{153.0mu}
          a(\la_{\ell_2}) c(\la_{M+1}, \la_{\ell_2})
	  \Bigr[ \prod_{\substack{k = 1 \\ k \ne \ell_1, \ell_2}}^{M+1}
	         \frac{1}{b(\la_k, \la_{\ell_2})} \Bigl] \<0|
		 \prod_{\substack{k = 1 \\ k \ne \ell_1, \ell_2}}^{M+1}
		 C(\la_k) \epp
\end{align}
These formula prove (\ref{genact}) for $m = 1$. They can also be used
in the induction step from $n \ge 1$ to $n + 1$.

In order to calculate the algebraic Bethe ansatz expression for the
finite Trotter number approximant $D_N$, equation (\ref{defdn}),
we have to set $M = N/2$ in (\ref{genact}), then insert the solution
$\{\la_j\}_{j=1}^{N/2}$ of the Bethe ansatz equation which belongs to
the leading eigenvalue, then multiply by $\bigl|\{\la_j\}_{j=1}^{N/2}
\bigr\>$ from the right and finally divide by $\<\{\la\}|\{\la\}\> 
\prod_{j = 1}^m \La_0 (\x_j)$. The sums on the right hand side of the
resulting formula are naturally divided into a part from 1 to $M$ over
Bethe roots $\la_j$ and a part from $M + 1$ to $\g_j$ over
inhomogeneities~$\x_k$. The sums over the Bethe roots (and only
these) can be transformed into integrals over the canonical contour
${\cal C}$, since for any function $f(\om)$ holomorphic on and inside
${\cal C}$ the identity
\begin{equation} \label{sumtoint}
     \int_{\cal C} \frac{d\om \, f(\om) }{2\p \i (1 + \fa (\om))} 
        = \sum_{j = 1}^{N/2} \frac{f(\la_j)}{\fa' (\la_j)}
	= - \int_{\cal C}
	    \frac{d\om \, f(\om) }{2\p \i (1 + \faq (\om))}
\end{equation}
holds. If $f(\om)$ is only meromorphic additional contributions appear
due to the poles of $f(\om)$. In any case it is clear that all
sums over Bethe roots in the expression for $D_N$ may be transformed
into integrals over the contour ${\cal C}$ and that, after performing
this transformation, a sum over multiple integrals will remain.
These integrals will involve not more than $m$ integrations, and the
leading term involving precisely $m$ integrations will be generated
by the $m$-fold sum from 1 to $M$ over the Bethe roots,
\begin{multline} \label{leadterm1}
     \Th = \sum_{\ell_1 = 1}^M
           \sum_{\substack{\ell_2 = 1 \\ \ell_2 \ne \ell_1}}^M
	   \dots
	   \sum_{\substack{\ell_m = 1 \\ \ell_m \ne \ell_1, \dots,
	   \ell_{m-1}}}^M
	   \biggl[ \prod_{j=1}^{|\a^+|} a(\la_{\ell_{\g_j^+}})
	           c(\la_{M + \a_j^+}, \la_{\ell_{\g_j^+}})
	           \prod_{\substack{k = 1 \\ k \ne \ell_1, \dots,
	   	                   \ell_{\g_j^+}}}^{M + \a_j^+}
	  	  \frac{1}{b(\la_k, \la_{\ell_{\g_j^+}})} \biggr] \\
	   \biggl[ \prod_{j=1}^{m-|\a^+|} d(\la_{\ell_{\g_j^-}})
	           c(\la_{\ell_{\g_j^-}}, \la_{M + \be_j^-})
	           \prod_{\substack{k = 1 \\ k \ne \ell_1, \dots,
	                     \ell_{\g_j^-}}}^{M + \be_j^-}
	  	  \frac{1}{b(\la_{\ell_{\g_j^-}}, \la_k)} \biggr]
	   \frac{\<\{\x\} \cup \{\la^-\}|\{\la\}\>}
	        {\<\{\la\}|\{\la\}\>\prod_{j=1}^m \La_0 (\x_j)} \epc
\end{multline}
where $\{\la^-\}$ is the complement of $\{\la^+\}$ in $\{\la\} =
\{\la_j\}_{j=1}^{N/2}$, and $\{\la^+\} = \{\la_{\ell_1}, \dots,
\la_{\ell_m}\}$. Inserting (\ref{ratio}) into the right hand side of
(\ref{leadterm1}) and using the Bethe ansatz equations (see
\cite{GKS04a}) we arrive at
\begin{multline}
     \Th = \sum_{\ell_1, \dots, \ell_m = 1}^M
	   \frac{\det( - G(\la_{\ell_j}, \x_k))}
	        {\prod_{1 \le j < k \le m} \sh(\x_k - \x_j)
		 \sh( \la_{\ell_k} - \la_{\ell_j} - \h)} \\
	   \biggl[
	      \prod_{k=1}^{|\a^+|} \frac{1}{\fa' (\la_{\ell_k})}
           \biggr]
           \biggl[
	   \prod_{j=1}^{|\a^+|}
	     \prod_{k=1}^{\a_j^+ - 1}
	     \sh(\la_{\ell_{|\a^+| - j + 1}} - \x_k - \h)
	     \prod_{k = \a_j^+ + 1}^m
	     \sh(\la_{\ell_{|\a^+| - j + 1}} - \x_k)
           \biggr] \\
           \biggl[
	      \prod_{k = |\a^+| + 1}^{m} \frac{-1}{\fa' (\la_{\ell_k})}
           \biggr]
           \biggl[
           \prod_{j = 1}^{m - |\a^+|}
	     \prod_{k=1}^{\be_j^- - 1}
	     \sh(\la_{\ell_{|\a^+| + j}} - \x_k + \h)
	     \prod_{k = \be_j^- + 1}^m
	     \sh(\la_{\ell_{|\a^+| + j}} - \x_k)
           \biggr] \epp
\end{multline}
Using (\ref{sumtoint}) this expression turns into (\ref{densint}) plus
a sum over terms involving less than $m$ integrals. In other words,
the right hand side of (\ref{densint}) is the unique leading term
as described above. Thus, our conjecture (\ref{densint}) means that
the subleading terms mutually cancel each other. As mentioned above
we verified this statement with a number of examples.
\section{Discussion}
We have presented new integral (\ref{densinthom}) formulae for the
density matrix of the XXZ chain and for its inhomogeneous
generalization (\ref{densint}). These formulae have been verified
for the general matrix element for small $m$ and for the special
case of the emptiness formation probabilities for all $m$ (note that
$\< {e_1}_1^1 \dots {e_m}_1^1 \>_{T,h}$ and $\< {e_1}_2^2 \dots
{e_m}_2^2 \>_{T,h}$ are different if $h \ne 0$). We also outlined two
important steps of the proof of the general formula on which we are
working now.
\enlargethispage{\baselineskip}

We think that our formulae have a great potential for applications in
the calculation of finite temperature static correlation functions
of the XXZ chain. We believe that the density matrix elements may
efficiently be summed up \cite{KMST04cpp}. We hope that for short
segments the multiple integrals may be reduced to single integrals
as e.g. in \cite{KSTS03}. We have started to evaluate some of
the integrals numerically. Last but not least we are very curious
if the analysis of the zero temperature inhomogeneous case as developed
in \cite{BJMST04} carries over to finite temperatures.

\enlargethispage{\baselineskip}
{\bf Acknowledgement.}
The authors would like to thank H. E. Boos, M. Bortz and
N. P. Hasenclever for helpful discussions. This work was supported by
the Deutsche Forschungsgemeinschaft under grant number Go 825/4-1.

\bibliographystyle{amsplain}
\bibliography{hub}

\end{document}